\documentclass{./aa}
\usepackage{amssymb}
\usepackage{graphics}
\usepackage[dvips]{epsfig}
\usepackage{natbib}
\bibpunct{(}{)}{;}{a}{}{,}

\newcommand{\rb}[1]{\raisebox{1.5ex}[-1.5ex]{#1}}
\newcommand{\decsec}[2]{$#1\mbox{$''\mskip-7.6mu.\,$}#2$}
\newcommand{\decsecmm}[2]{#1\mbox{$''\mskip-7.6mu.\,$}#2}

\begin{document}     

\title{The stellar content of the Hamburg/ESO survey\thanks{Based on
    observations collected at the European Southern Observatory, Chile
    (Proposal IDs 145.B-0009 and 63.L-0148). Table B.1 is only available in
    electronic form at the CDS via anonymous ftp to cdsarc.u-strasbg.fr
    (130.79.125.5) or via
    http://cdsweg.u-strasbg.fr/Abstract.html.  }  }
\subtitle{II. A large, homogeneously-selected sample of high latitude carbon
  stars}

\author{N. Christlieb\inst{1}
\and P.J. Green\inst{2}  
\and L. Wisotzki\inst{3}
\and D. Reimers\inst{1}
}

\institute{Hamburger Sternwarte, Universit\"at Hamburg, Gojenbergsweg 112,
  D-21029 Hamburg, Germany\\
  \email{nchristlieb/dreimers@hs.uni-hamburg.de}
\and Harvard-Smithsonian Center for Astrophysics, 60 Garden Street, 
  Cambridge, MA 02140, USA\\
  \email{pgreen@cfa.harvard.edu}
\and Institut f\"ur Physik, Universit\"at Potsdam, Am Neuen Palais 10, 
  D-14469 Potsdam, Germany\\
  \email{lutz@astro.physik.uni-potsdam.de}
}

\offprints{nchristlieb@hs.uni-hamburg.de}
\date{Received 26-April-2001; accepted 11-June-2001}
\titlerunning{Carbon Stars from the Hamburg/ESO Survey}
\authorrunning{Christlieb et al.}

\abstract{
  We present a sample of 403 faint high latitude carbon (FHLC) stars selected
  from the digitized objective prism plates of the Hamburg/ESO Survey (HES).
  Because of the $\sim 15$\,{\AA} spectral resolution and high signal-to-noise
  ratio of the HES prism spectra, our automated procedure based on the
  detection of C$_2$ and CN molecular bands permits high-confidence
  identification of carbon stars without the need for follow-up spectroscopy.
  From a set of 329 plates (87\,\% of the survey), covering $6\,400$\,deg$^2$
  to a magnitude limit of $V\sim16.5$, we analyze the selection efficiency and
  effective surface area of the HES FHLC survey to date.  The surface density
  of FHLC stars that we detect ($0.072\pm 0.005$\,deg$^{-2}$) is 2--4 times
  higher than that of previous objective prism and CCD surveys at high
  galactic latitude, even though those surveys claimed a limiting magnitude up
  to $1.5$ magnitudes fainter. This attests to the highest selection
  sensitivity yet achieved for these types of stars.  \keywords{Stars:carbon
    -- Surveys -- Galaxy:halo} }

\maketitle

\section{Introduction}

Models of the chemical and dynamical properties of the Galactic spheroid (the
`halo') are still rather weakly constrained.  In the grand scheme, did a
monolithic protogalaxy undergo rapid collapse and enrichment
\citep{Eggenetal:1962}, or did many smaller dwarf galaxies merge together
\citep{Searle/Zinn:1978}? Both processes probably contribute, since there is
solid recent evidence of ongoing mergers
\citep{Ibataetal:1994,Majewskietal:2000}. Stars and gas that are tidally
stripped from accreting dwarf galaxies remain aligned with the orbit of the
satellite for timescales comparable to the age of the Galaxy. Thus, a number
of tidal streams exist today whose phase-space signature can constrain the
stripping and merging events that contribute to the present-day galactic halo
\citep{Johnstonetal:1999}.

An important goal of astronomy in this century is to measure and model the
potential of the Milky Way using halo stars as tracers.  To simultaneously
disentangle the remnants of disrupted satellites requires full knowledge of the
angular positions, proper motions, radial velocities, and distances of a large
number of such stars.  But first, a large sample of distant halo stars must be
amassed.  Intrinsically bright stars visible to large galactocentric distances
($10$--$100$~kpc) provide the best opportunity. Because they are readily
recognizable from their strong C$_2$ and CN absorption bands, and because they
were thought to be giants without exception, faint C stars have been sought as
excellent tracers of the outer halo.

Faint high galactic latitude carbon (FHLC) stars have been sought in prior
objective prism surveys
\citep[e.g.,][]{Sanduleak/Pesch:1988,MacAlpine/Lewis:1978} and in the CCD
survey of \cite{Greenetal:1994}.  Objective-prism photography with wide-field
Schmidt telescopes has yielded low-dispersion spectra for thousands of objects
over substantial portions of the sky, but not a large number of carbon stars.
Fewer than 1\% of the $6\,000$ stars in Stephenson's (1989) catalogue are the
faint, high-latitude carbon (FHLC) stars ($V>13,~|b|>40^{\circ}$) most useful
as dynamical probes of the outer halo.  The two most prolific sources of
published FHLC stars, the Case low-dispersion survey
\citep[CLS;][]{Sanduleak/Pesch:1988} and the University of Michigan -- Cerro
Tololo survey \citep[UM;][]{MacAlpine/Williams:1981} appear to probe to about
$V=16$ and have provided about 30 FHLC stars.  Emission-line objects, not FHLC
stars, were the primary goal of these photographic surveys, and known FHLC
stars were not examined to help predefine selection criteria or estimate
completeness. The surface density of FHLC stars from objective-prism surveys
is low, about one per 50 deg$^2$ to $V\approx16$.  At high galactic latitudes,
mostly warm carbon stars are found -- CH stars, and possibly some R stars.
However, color selection of very red stars at high latitude also reveals a
small number (one per 200 deg$^2$ to $R\sim 16$) of classical intermediate age
AGB carbon (AGBC) stars \citep{Totten/Irwin:1998}. \cite{Margonetal:2000}
recently reported the discovery of more than 30 new FHLC stars in the
commissioning data of the Sloan Digitized Sky Survey (SDSS), which may
eventually provide the majority of known FHLC stars.

\begin{figure}[tbp]
  \begin{center}
    \epsfig{file= 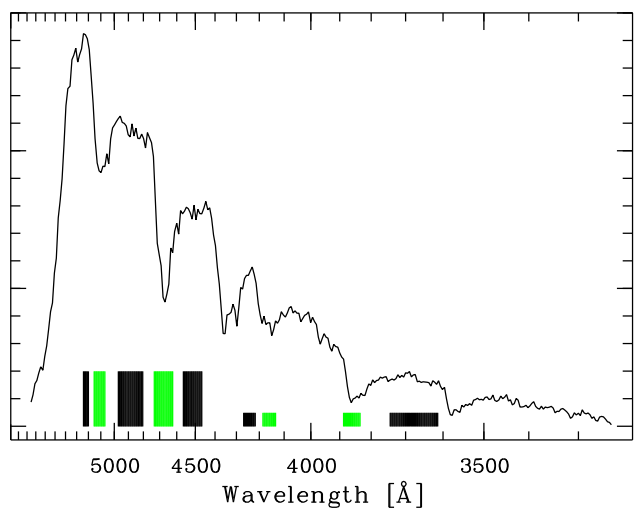, clip=, width=8.5cm,
      bbllx=138, bblly=572, bburx=371, bbury=758}
    \caption{\label{fig:Cbands}
      HES objective prism spectrum of the R-type carbon star CGCS~2954
      \citep{Stephenson:1989}, illustrating the positions of
      continuum (black) and band (grey) bandpasses defining the C$_2$ (high
      boxes) and CN (flat boxes) line indices used for selection in the
      HES. The abscissa is density above diffuse sky background in
      arbitrary units. Note that wavelength is \emph{decreasing\/} towards
      the right. The sharp drop of the spectra at $\lambda\sim 5400$\,{\AA}
      is due to the IIIa-J emulsion sensitivity cutoff.}
  \end{center}
\end{figure}

In this paper, we describe our use of the Hamburg/ESO survey
\citep[HES;][]{hespaperI,heshighlights,hespaperIII} to greatly augment the
number of known FHLC stars. The HES is an objective-prism survey designed to
select bright ($12.5 \gtrsim B_J \gtrsim 17.5$) quasars in the southern
extragalactic sky ($\delta<+2.5^\circ$; $|b|\gtrsim 30^\circ$). It is based on
IIIa-J plates taken with the 1\,m ESO Schmidt telescope and its
4$^{\circ}$ prism, yielding a wavelength range of $3200\,\mbox{\AA} < \lambda
< 5200\,\mbox{\AA}$ and a seeing-limited spectral resolution of typically
15\,{\AA} at H$\gamma$. This resolution makes possible the identification of
carbon stars with high confidence without follow-up slit spectroscopy, based
on their strong C$_2$ and CN molecular bands (cf. Fig. \ref{fig:Cbands}).

Since carbon can reach the surface of an isolated star only in late
evolutionary stages, it has long been assumed that all carbon (C) stars are
giants.  Due to their high luminosity ($M_R\sim -3.5$), it is possible to
detect the red AGBC stars at large distances: \cite{Breweretal:1996} have
identified C stars even in the local group galaxy M31.  The more typical FHLC
stars such as CH giants, with $0< M_V < -2.5$, can be detected to $\sim
60$\,kpc in sensitive photographic surveys.  However, the long-held assumption
that all C stars are giants has fallen.  Trigonometric parallax measurements
for the carbon star \object{G77--61} \citep{Dahnetal:1977} showed that this
star lies close to the main sequence ($M_V\sim+10$). For years, the dwarf
carbon (dC) stars phenomenon was assumed to be extremely rare until many new
dCs were discovered in the early 1990s
\citep{Greenetal:1991,Greenetal:1992,Warrenetal:1993,Heberetal:1993,Liebertetal:1994}.
Discovery of so many dCs, and the remarkable similarity of their spectra to
those of C giants means that care must be taken to distinguish dwarfs from
giants in FHLC star samples intended for distant halo studies
\citep{Greenetal:1992}.  We are therefore undertaking a two-part
investigation.  In the current paper, we describe our automated selection of C
stars in the HES, and we present a large, uniformly selected, and flux-limited
sample of FHLC stars. We complement this sample in upcoming work with recent
epoch astrometry, to measure proper motions for as many objects as possible, and
thereby separate the dCs from C giants.

We note that carbon-enhanced, metal-poor stars may be among the FHLC stars
presented here. It was recognized by several authors that the fraction of such
stars among metal-poor stars rises with decreasing metallicity, reaching
$\sim 25$\,\% for stars with $\mbox{[Fe/H]}<-3.0$, and that carbon overabundances
are as high as $\mbox{[C/Fe]}=+2.0$\,dex
\citep[e.g.,][]{Norrisetal:1997,Rossietal:1999}. $^{12}$C/$^{13}$C isotope
measurements of a larger sample of such stars would help to identify the
carbon production site(s) at work.

\section{Carbon Star Selection}\label{selection}

The full HES database consists of $\sim 10$ million extracted, wavelength
calibrated spectra. The input catalog for extraction of objective-prism
spectra is generated by using the Digitized Sky Survey I (DSS~I). An
astrometric transformation between DSS~I plates and HES plates yields, for
each object in the input catalog, the location of its spectrum on the relevant
HES plate, and provides a wavelength calibration zero point
\citep{hespaperIII}.

\begin{figure}[tbp]
  \begin{center}
    \epsfig{file= 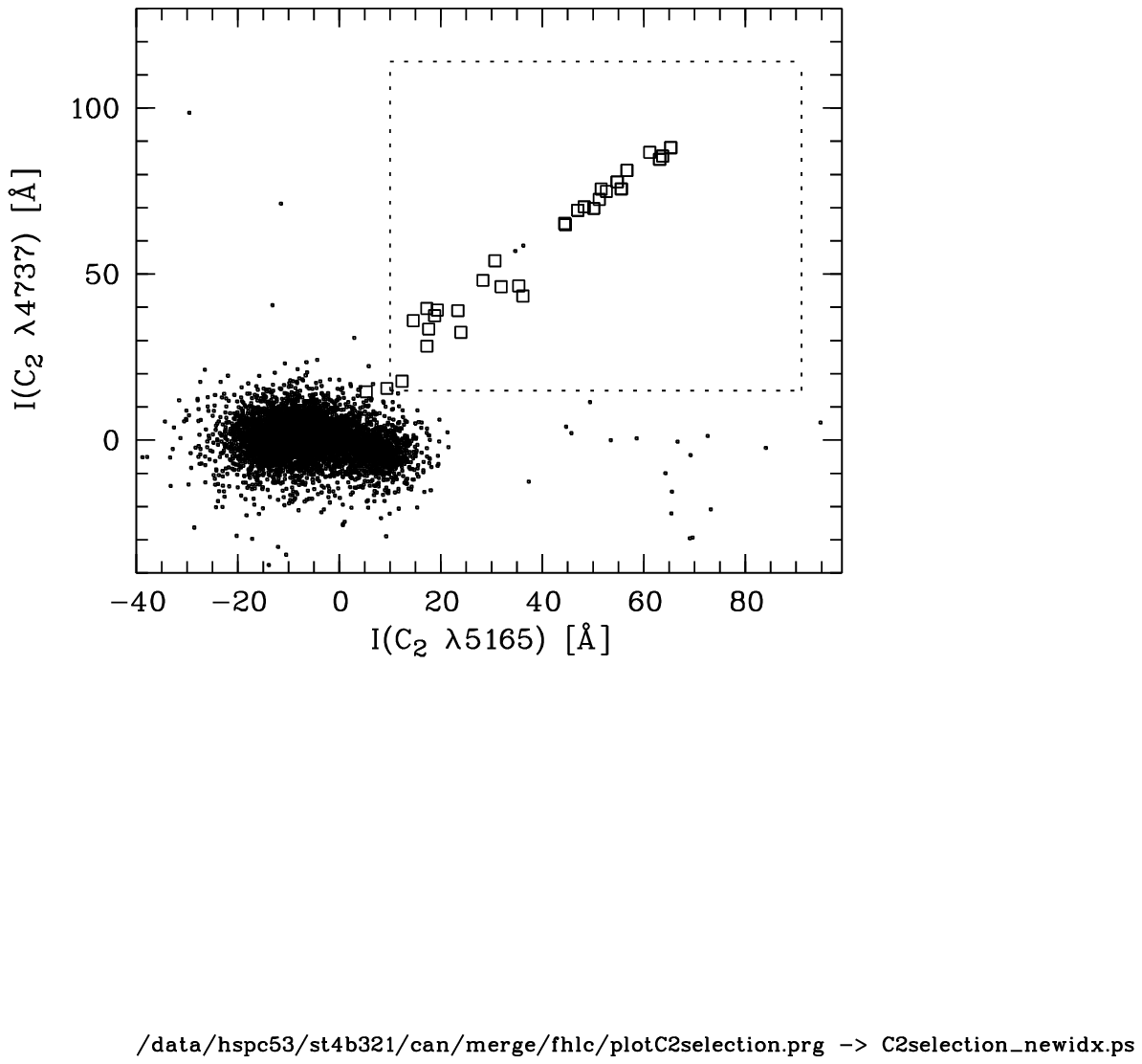, clip=, width=8.5cm,
      bbllx=101, bblly=572, bburx=356, bbury=773}
    \caption{\label{C2selection}
      Selection of carbon stars in the $I\,(\mbox{C}_2\;\lambda\,5165)$ versus
      $I\,(\mbox{C}_2\;\lambda\,4737)$ plane.  Band strengths are measured by
      line indices.  `$\cdot$' -- all spectra on a randomly
      chosen HES plate, `$\square$' -- test sample of known FHLC stars present
      on HES plates (see Tab. \ref{tab:seltest}), dashed box -- selection
      region. Spectra in which only {\em one\/} high C$_2$ band index value
      was measured suffer either from an overlapping spectrum, or from a
      plate artifact.  The selection in the $I\,(\mbox{CN}\;\lambda\,4216)$
      versus $I\,(\mbox{CN}\;\lambda\,3883)$ plane is done analogously. The
      two test sample objects outside the selection box are CGCS~525 and
      CGCS~3180. They are selected by CN band indices (see Tab.
      \ref{tab:seltest}). }
  \end{center}
\end{figure}

Carbon stars can be identified in the HES data base by their strong C$_2$ and
CN bands.  We select carbon star candidates when the mean signal-to-noise
ratio ($S/N$) in the relevant wavelength range is $>5$ per pixel and both of
the C$_2$ bands $\lambda\lambda\,5165$, $4737$, or both of the CN bands
$\lambda\lambda\,4216$, $3883$ are stronger than a selection threshold. Band
strengths are measured by means of line indices -- ratios of the mean
photographic densities in the carbon molecular absorption features and the
continuum bandpasses shown in Fig. \ref{fig:Cbands}, and listed in Tab.
\ref{tab:Cbands}. The use of {\em pairs\/} of indices prevents confusion with
plate artifacts, e.g., scratches.  It is very unlikely that two such artifacts
are present at the positions of two molecular bands. Selection boxes in the
$I\,(\mbox{C}_2\;\lambda\,5165)$ versus $I\,(\mbox{C}_2\;\lambda\,4737)$ and
$I\,(\mbox{CN}\;\lambda\,4216)$ versus $I\,(\mbox{CN}\;\lambda\,3883)$ planes
were chosen well-separated from the dense locus of ``normal'' stars (see Fig.
\ref{C2selection}). The selection criteria are listed in Tab.
\ref{tab:selectcrit}.

\begin{table}[tbp]
  \begin{flushleft}
    \caption{\label{tab:Cbands}
      Wavelengths of passbands used for computation of C band indices in
      the HES. `cont'=continuum; `flux'= feature passband.}
    \begin{tabular}{lcccc}\hline\hline
      \rule{0.0ex}{2.3ex} & \multicolumn{4}{c}{Use for band index}\\
      \rb{Passband} & C$_2$~5165 & C$_2$~4737 & CN~4216 & CN~3883\\\hline
      5190--5240\,{\AA} & cont &      &      &\rule{0.0ex}{2.3ex}\\
      5060--5150\,{\AA} & flux &      &      &\\
      4800--4970\,{\AA} & cont & cont &      &\\
      4620--4730\,{\AA} &      & flux &      &\\
      4460--4560\,{\AA} &      & cont &      &\\
      4210--4270\,{\AA} &      &      & cont &\\
      4130--4180\,{\AA} &      &      & flux &\\
      3830--3890\,{\AA} &      &      &      & flux\\
      3610--3740\,{\AA} &      &      &      & cont\\\hline\hline
    \end{tabular}    
  \end{flushleft}
\end{table} 

\begin{table}[tbp]
  \begin{flushleft}
    \caption{\label{tab:selectcrit} Carbon star selection criteria. The
      maximum allowed band index values correspond to an integrated density
      of zero in the feature passbands. That is, larger band indices can
      only be due to artifacts, e.g. scratches, causing photographic
      densities (above skybackground) $<0$. Stars are selected if both of
      their C$_2$ indices or both of their CN indices fall into the indicated
      ranges
      }
    \begin{tabular}{lc}\hline\hline
      Feature & Index range [{\AA}]  \rule{0.0ex}{2.3ex}\\\hline
      $\mbox{C}_2\;\lambda\,5165$ & [10,91]  \rule{0.0ex}{2.3ex}\\
      $\mbox{C}_2\;\lambda\,4737$ & [15,114] \\
      $\mbox{CN}\;\lambda\,4216$  & [2,56]   \\
      $\mbox{CN}\;\lambda\,3883$  & [13,55]   \\\hline\hline
    \end{tabular}    
  \end{flushleft}
\end{table} 

\begin{figure}[tbp]
  \begin{center}
    \leavevmode
    \epsfig{file=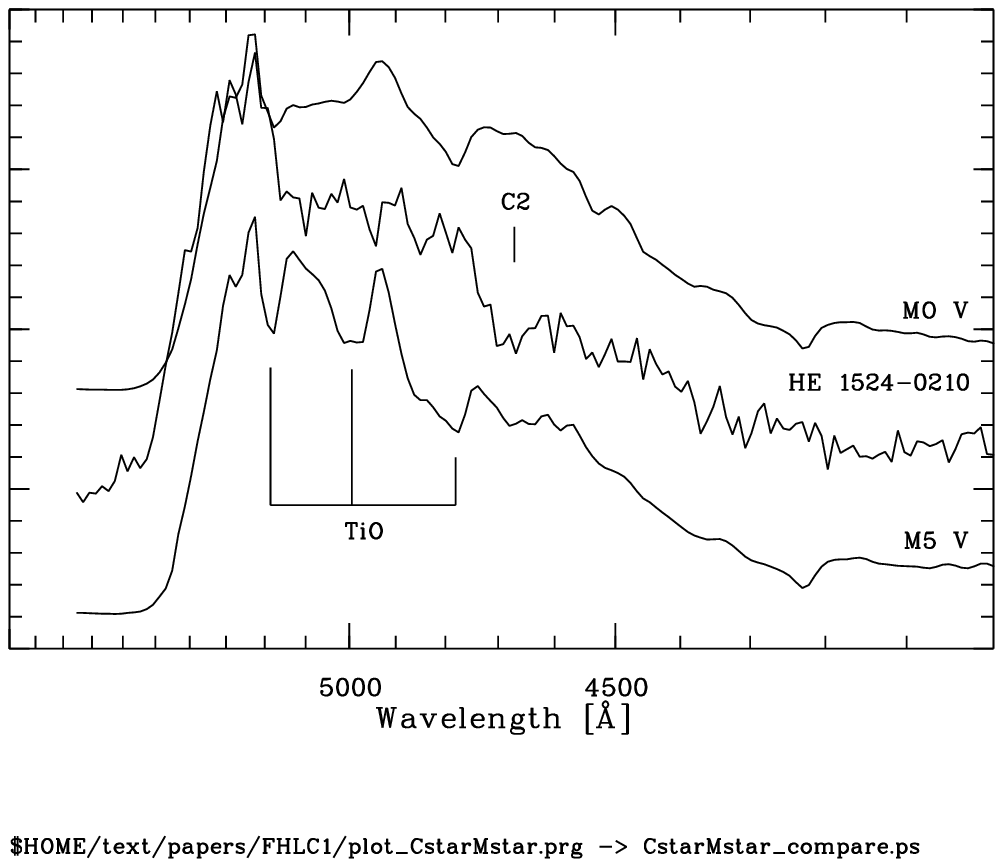, clip=, width=8.5cm,
      bbllx=80, bblly=426, bburx=370, bbury=641}
    \caption{
    \label{CstarMstar_compare} Comparison of HES spectra of the C star
    HE~1524$-$0210, exhibiting a weak C band only, with two M stars. The
    abscissa is the same as in Fig. \ref{fig:Cbands}.}  
  \end{center}
\end{figure}

Carbon stars can be distinguished reliably from other late type stars, e.g. M
or S stars, even if only weak C bands are present in their spectra (cf. Fig.
\ref{CstarMstar_compare}).
Other potential sample contaminators are white dwarfs of type DQ, which show
carbon molecular bands. However, since the latter usually have a much bluer
continuum (see Fig. \ref{Cstar_templates}), they can easily be recognized by
visual inspection of the spectra, and by their $U-B$ color.
\cite{McCook/Sion:1999} list 49 DQs, of which 30 have an available $U-B$
measurement. The average $U-B$ of those is $-0.58$, i.e., $\sim 1.5$\,mag away
from the average $U-B$ of the HES C star sample.  Our $U-B$ colors are
measured directly from the HES spectra with a mean accuracy of
$\sigma_{U-B}=0.09$\,mag \citep[][herafter Paper~I]{HESStarsI}.  The
average $U-B$ of HES C stars is $\sim 0.9$, more than 90\,\% have $U-B>0.5$,
and there is \emph{no} C star of $U-B<0$ in the HES sample. While 4 (i.e.,
13\,\%) of the 30 DQs with $U-B$ in \cite{McCook/Sion:1999} have $U-B>0.0$,
the pressure-broadened features of DQs are easily distinguished by visual
inspection of the carbon bands (see Fig. \ref{Cstar_templates}).

\begin{figure*}[tbp]
  \begin{center}
    \leavevmode
    \epsfig{file=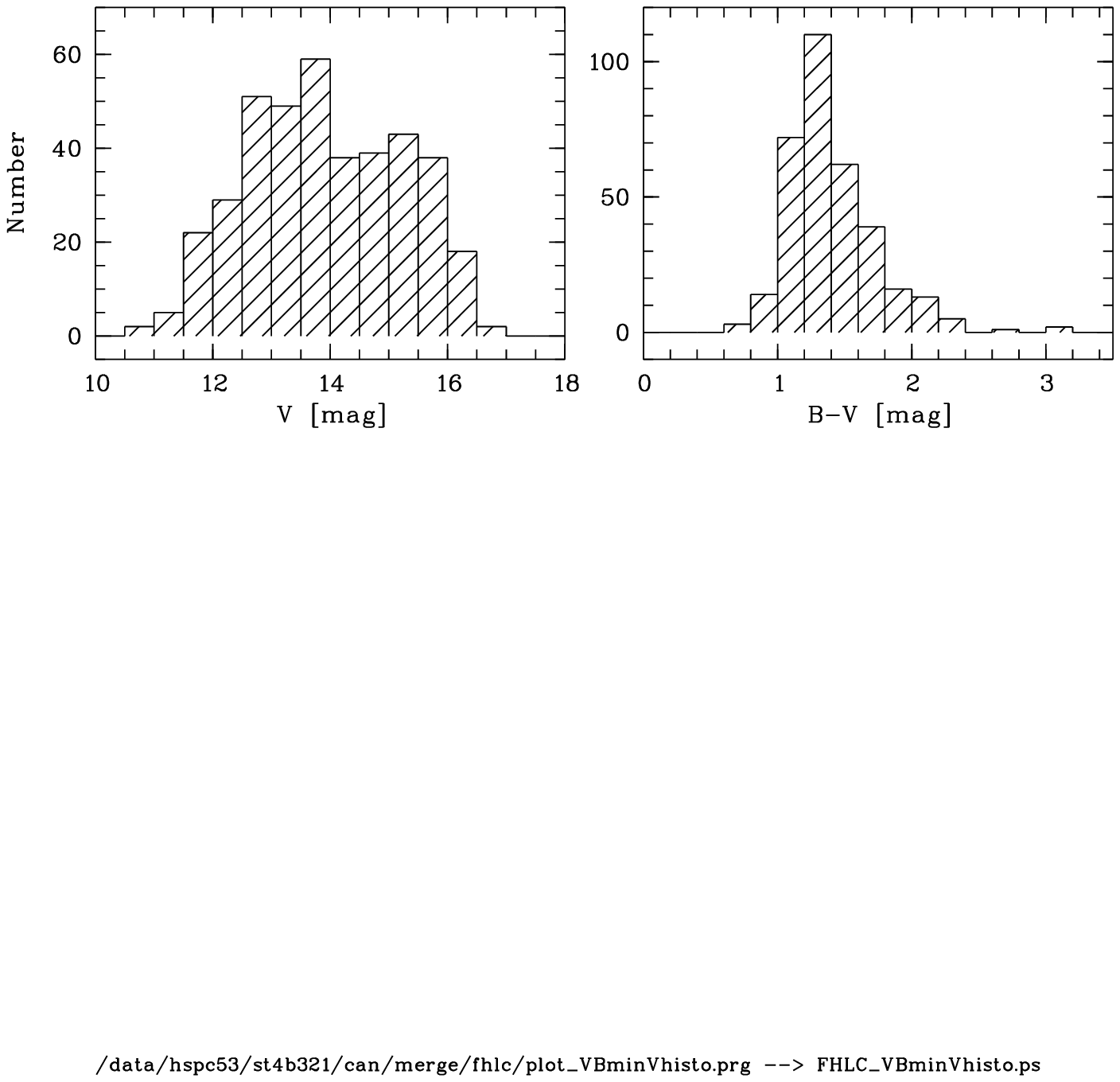, clip=, width=14cm,
      bbllx=77, bblly=425, bburx=484, bbury=584}
  \end{center}
  \caption{
    \label{FHLC_VBminVhisto} $V$ magnitude and $B-V$ distribution of the HES FHLC
    sample. $B-V$ was derived from HES spectra with the procedures
    described in Paper~I.}  
\end{figure*}

With a rough estimate of their surface density, we can quantify an
upper limit for the contamination of the HES C star sample by ``red''
($U-B>0.0$) DQs. First of all, we have to take into account that the
ratio of northern hemisphere to southern hemisphere DQs is unbalanced
in \cite{McCook/Sion:1999}, as much as the \emph{total} catalog
is. This is because the southern hemisphere so far has been surveyed
less extensively for white dwarfs. Assuming that the northern
hemisphere sample of DQs is complete, we derive a surface density of 9
DQs brighter than $V=16.5$ in $20\,000$\,deg$^{2}$, i.e. $4.5\cdot
10^{-4}$\,deg$^{-2}$. Hence, the surface density of $U-B>0.0$ DQs is
$5.9\cdot 10^{-5}$\,deg$^{-2}$, and we expect $0.44$ DQs to be present
on all 329 plates currently used for the exploitation of the stellar
content of the HES.  Therefore, even if we assume that the sample of
DQs known so far is incomplete by a factor of 2, we statistically
expect less than 1 DQ to be present in the HES C star sample.

\begin{figure}[tbp]
  \begin{center}
    \leavevmode
    \epsfig{file=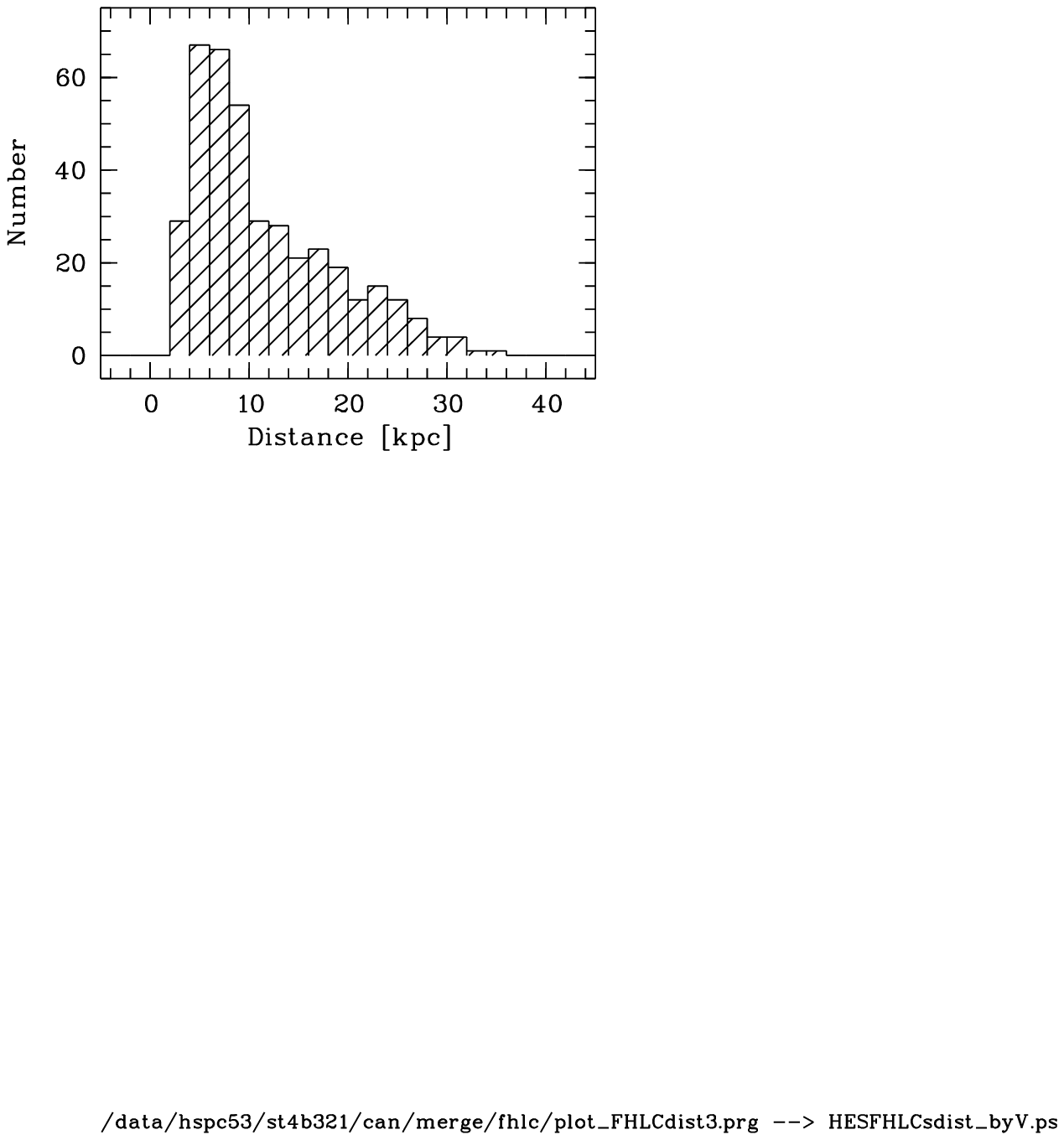, clip=, width=7cm,
      bbllx=77, bblly=425, bburx=286, bbury=584}
  \end{center}
  \caption{
    \label{HESFHLCsdist} Distance distribution of the 393 HES FHLCs
    with available $V$ magnitudes, assuming that they are all giants with
    $M_V=-1$\,mag.}  
\end{figure}

On the 329 HES plates (effective area $6\,400$\,deg$^2$) we found 403 FHLCs.  90
of them were selected by C$_2$ band indices only, 171 by CN band indices only,
and 144 by C$_2$ and CN indices. The $V$ and $B-V$ distributions are displayed
in Fig. \ref{FHLC_VBminVhisto}. The faintest objects have $V\sim 16.5$, and
the most distant objects reach $\sim 35$\,kpc (cf. Fig. \ref{HESFHLCsdist}),
assuming they are all giants with $M_V=-1$\,mag.

\section{Testing the Automated Selection}

We tested the automated selection extensively and by various methods.  In
Sect. \ref{sect:sel_eff} we investigate the selection efficiency. In Sect.
\ref{sect:dC_selfunct} we derive plate-by-plate selection probabilities for halo
dCs on HES plates by simulations. The results of tests with ``real'' objects
are given in Sect. \ref{Cstar_testobjects}.

\subsection{Selection Efficiency}\label{sect:sel_eff}

An important criterion for the evaluation of the quality of a selection
algorithm is the \emph{selection efficiency}, i.e. the fraction of desired
stars in the raw candidate sample. Tab. \ref{tab:sel_eff} summarizes the
results.
\begin{table}[tbp]
    \caption{\label{tab:sel_eff}
      Selection efficiency for C stars in the HES. UNID=\emph{probable} C
      stars with weak C bands, OVL=overlapping spectra, ART=artifacts,
      NOIS=very noisy spectra, SAT=saturated spectra. The raw candidate
      reduction factor is the factor by which the selection
      algorithm reduces the \emph{total} set of $3\,437\,630$ overlap-free
      HES spectra with $S/N>5$ present on 329 HES plates.
      }
  \begin{flushleft}
    \begin{tabular}{lrr}\hline\hline
      Raw candidate  & \rule{0.0ex}{2.3ex}\\[-0.3ex]
    reduction factor & \rb{1/2900} \\[0.5ex]
    C stars & 31.6\,\%  \\
    UNID    &  7.0\,\%  \\
    OVL     & 29.2\,\%  \\
    ART     &  8.7\,\%  \\
    NOIS    &  3.8\,\%   \\
    SAT     & 15.6\,\%  \\\hline\hline
    \end{tabular}    
  \end{flushleft}
\end{table}
Our selection is very efficient. The low fraction of artifacts demonstrates
that the usage of \emph{pairs} of C$_2$ bands and CN bands indeed very
reliably excludes artifacts from selection. However, a considerable number of
overlapping spectra (overlaps) are selected. Overlaps are detected by an
automatic overlap detection algorithm in the HES, using the direct plate data
of the DSS~I. It appears that our carbon star selection technique is
very sensitive in finding the small number of overlaps not detected by the
automatic algorithm.

\subsection{Decrease of Selection Probability for Halo
  dCs}\label{sect:dC_selfunct}

In the HES, some care must be taken when objects with large proper motion are
selected.  This is because the input catalog for extraction of objective prism
spectra is generated by using the DSS~I. The dispersion direction of the HES
spectra is along declination. Therefore, large proper motions and/or large
epoch differences between HES and DSS~I plates (13.5 years on average) may
result in an offset of the wavelength calibration zero point, leading to
smaller C band values, and/or non-detection of objects in the HES, if
$\mu_{\alpha}\Delta\,t_{\mbox{\scriptsize HES-DSS~I}}\gtrsim 4''$, i.e., $>$ 3
pixels. Offsets of $\mu_{\alpha}\Delta\,t_{\mbox{\scriptsize HES-DSS~I}}< 4''$
can be recovered by the spectrum extraction algorithm.  Note however that
proper motions of a typical halo object ($<\!\!u\!\!>=<\!\!w\!\!>=0$\,km/s;
$<\!\!v\!\!>\sim 200$\,km/s) result in $\sim 2\times$ larger offsets along
declination than in R.A., since the galactic plane is tilted by $62.6^{\circ}$
with respect to the equatorial coordinate system.

In order to estimate how many dCs are expected to be missed in our survey due
to the epoch difference problem, we carried out a simulation study in which
the plate-by-plate selection function for halo dCs was determined. The simulation
is similar to that described in \cite{Greenetal:1992}. We employ a sample of
simulated dCs with halo kinematics, as given by \cite{Norris:1986}. For the
solar neighborhood, he gives
\begin{equation}
  \label{eq:vrot}
  v_{\mbox{\scriptsize rot}} = 37\pm 10\,\mbox{km/s}
  \Longleftrightarrow <\!\!v\!\!> = -187\,\mbox{km/s},
\end{equation}
and he determined the velocity ellipsoid to be
\begin{eqnarray}
  \sigma_u&=&131\pm 6\,\mbox{km/s}\label{eq:sigma_u}\\
  \sigma_v&=&106\pm 6\,\mbox{km/s}\label{eq:sigma_v}\\
  \sigma_w&=& 85\pm 6\,\mbox{km/s}\label{eq:sigma_w}.
\end{eqnarray}
In each simulation we constructed 100 random velocity vectors $(u,v,w)$, with
components following Gaussian distributions according to the above parameters.
These velocity vectors were each applied to stars located at the center of the
plate under investigation, and converted to proper motions assuming distances
$d$.  These were computed from the apparent $V$ magnitude distribution of a
sample of 86 C stars without significant p.m. (see Fig.
\ref{FHLC_VBminVhisto}), and assuming $M_V=+10$ for dwarf carbon stars. This
yields $86\cdot 100=8\,600$ simulated stars.  We compute the position of the
star after the time $\Delta t_{\mbox{\scriptsize HES-DSS~I}}$, the epoch
difference between DSS~I and HES plate, and derive proper motions
$\mu_\alpha$, $\mu_\delta$ from the position differences.  We then select the
subsample of the 8\,600 stars with $\mu_{\alpha}\Delta\,t_{\mbox{\scriptsize
    HES-DSS~I}}< 4''$.

As a test sample for an investigation of the dependence of the selection
probability on $\mu_\delta\Delta t$, we used a sample of 78 C stars from 44
HES plates \emph{without} significant p.m., as measured in an follow-up
campaign carried out in April at ESO, using the Wide Field Imager
attached to the MPG/ESO 2.2\,m telescope (Christlieb et al. 2001, in
preparation). The stars were shifted in 1 pixel ($=\decsecmm{1}{35}$) steps
through the range $-700\,\mu\,\mbox{m}<x<+700\,\mu\,\mbox{m}$, corresponding
to $-\decsecmm{47}{25}<\mu_\delta\Delta t<\decsecmm{47}{25}$. At each shift
step, the selection algorithms were applied.

By applying the selection probability (a function of $\mu_\delta\Delta t$) to
the subsample of the 8\,600 stars with
$\mu_{\alpha}\Delta\,t_{\mbox{\scriptsize HES-DSS~I}}< 4''$, we determine the
fraction of stars which would be detected in the HES \emph{and} selected by
our selection algorithm. On the 329 stellar HES plates, $21.4$\,\% of the
simulated halo dCs are detected and selected.

\cite{Greenetal:1992} found that 13\,\% of their C stars are dwarfs. Applying
this estimate to our sample, and taking into account that we find only $\sim
20$\,\% of the dCs detectable on the HES plates, we estimate that 10--15 out
of our 403 FHLCs are dCs. However, since our sample is biased to low-p.m., it
is likely that not all of these can be \emph{proven} to be dCs by their large
transverse velocity. Based on our simulations we estimate that additional
$\sim 40$ dCs would be detectable on the HES plates, but are currently missed
due to the epoch difference problem. This incompleteness will be addressed in
a later paper focusing on dC stars in the HES.

\subsection{Tests with Known C Stars}\label{Cstar_testobjects}

We also compiled a test sample of known dwarf and giant C stars present on HES
plates (see Tab. \ref{tab:seltest}). We took all three dCs in the southern
hemisphere listed by \cite{Deutsch:1994}, i.e.  LHS~1075, G77--61, and KA~2.
The (possible) dCs of \cite{Warrenetal:1993}, having $B_J>20$, unfortunately
are by far too faint to be detectable on HES plates.  Cross-identification
with the C star lists of \cite{Slettebaketal:1969a}, \cite{Stephenson:1989},
\cite{Bothunetal:1991}, and \cite{Totten/Irwin:1998}, yielded 21 stars.
Another 6 spectra were produced from slit spectra with the procedures
described in Paper~I.

\begin{table*}[tbp]
    \caption{\label{tab:seltest} Test
      sample of dwarf and giant C stars present on HES plates. Sources:
      BEM91=\cite{Bothunetal:1991}, D94=\cite{Deutsch:1994},
      S89=\cite{Stephenson:1989},
      SKB69=\cite{Slettebaketal:1969a}, TI98=\cite{Totten/Irwin:1998}. Stars
      marked with TI98 have been recently reported by \cite{Tottenetal:2000}
      to have no significant p.m. Therefore, we list them with
      ($\mu_{\alpha}\Delta\,t,\mu_{\delta}\Delta\,t)=(0,0)$. KA~2 was not
      selected by CN bands, using its real spectrum, but it \emph{was}
      selected in the course of our simulations. This is because the $S/N$ at
      the position of the CN bands is too low in the former spectrum, and
      effectively infinite in the simulated spectrum}
  \begin{center}
    \begin{tabular}{llccrrcccl}\hline\hline
      & & & & & & \multicolumn{3}{c}{Selected by}\\
      \rb{Name} & \rb{HE Name} & \rb{$B_J$} & \rb{$B-V$} &
      \rb{$\mu_{\alpha}\Delta\,t$} & \rb{$\mu_{\delta}\Delta\,t$} &
          C$_2$ & CN & All & Source\rule{0.0ex}{2.3ex}\\\hline
 CGCS~39      & HE 0017$+$0055 &(sat.)&     &     &     & 1 & 1 &  1 & S89\rule{0.0ex}{2.3ex}\\
 SKB~2        & HE 0039$-$2635 & 13.1 & 1.1 &     &     & 1 & 1 &  1 & SKB69 \\
 BEM91~23     & HE 0100$-$1619 & 15.9 & 1.5 &     &     & 1 & 0 &  1 & BEM91 \\
 CGCS~177     & HE 0106$-$2837 & 13.8 & 2.1 &     &     & 1 & 0 &  1 & S89 \\
 SKB~5        & HE 0111$-$1346 & 13.3 & 1.4 &     &     & 1 & 1 &  1 & SKB69 \\
 0207$-$0211  & HE 0207$-$0211 & 15.5 & 2.2 & 0.0 & 0.0 & 1 & 0 &  1 & TI98\\
 BEM91~08     & HE 0228$-$0256 & 16.2 & 2.0 &     &     & 1 & 0 &  1 & BEM91 \\
 CGCS~525     & HE 0330$-$2815 & 13.8 & 1.5 &     &     & 0 & 1 &  1 & S89 \\
 CGCS~935     & HE 0521$-$3425 & 13.0 & 1.3 &     &     & 1 & 1 &  1 & S89 \\
 0915$-$0327  & HE 0915$-$0327 & 14.5 & 2.3 & 0.0 & 0.0 & 1 & 0 &  1 & TI98\\
 1019$-$1136  & HE 1019$-$1136 & 15.2 & 1.8 & 0.0 & 0.0 & 1 & 0 &  1 & TI98\\
 CGCS~2954    & HE 1104$-$0957 &(sat.)&     &     &     & 1 & 1 &  1 & S89 \\
 CGCS~3180    & HE 1207$-$3156 & 12.8 & 1.2 &     &     & 0 & 1 &  1 & S89 \\
 CGCS~3274    & HE 1238$-$0836 &(sat.)&     &     &     & 1 & 1 &  1 & S89 \\
 1254$-$1130  & HE 1254$-$1130 & 16.1 & 2.2 & 0.0 & 0.0 & 1 & 0 &  1 & TI98\\
 1339$-$0700  & HE 1339$-$0700 & 15.0 & 1.7 & 0.0 & 0.0 & 1 & 0 &  1 & TI98\\
 1442$-$0058  & HE 1442$-$0058 & 17.8 & 2.2 & 0.0 & 0.0 & 1 & 0 &  1 & TI98\\
 CGCS~5435    & HE 2144$-$1832 & 12.6 & 1.4 &     &     & 0 & 1 &  1 & S89 \\
 CGCS~5549    & HE 2200$-$1652 & 12.3 & 0.9 &     &     & 1 & 1 &  1 & S89 \\
 2213$-$0017  & HE 2213$-$0017 & 16.4 & 2.4 & 0.0 & 0.0 & 1 & 0 &  1 & TI98\\
 2225$-$1401  & HE 2225$-$1401 & 16.5 & 2.9 & 0.0 & 0.0 & 1 & 0 &  1 & TI98\\
 CLS~50       &                &      &     & 0.0 & 0.0 & 1 & 0 & -- & Simul.\\   
 CLS~31       &                &      &     & 0.0 & 0.0 & 1 & 1 & -- & Simul.\\   
 CLS~54       &                &      &     & 0.0 & 0.0 & 1 & 1 & -- & Simul.\\   
 KA~2         &                &      &     & 0.0 & 0.0 & 1 & 1 & -- & Simul.\\   
 B1509$-$0902 &                &      &     & 0.0 & 0.0 & 1 & 1 & -- & Simul.\\   
 UM~515       &                &      &     & 0.0 & 0.0 & 1 & 0 & -- & Simul.\\   
%-----------------------------------------------------------------------%
 LHS~1075   & HE 0023$-$1935 & 16.1 & 1.4 & $-$\decsec{0}{24} & $-$\decsec{10}{0} 
                                                    & 0 & 0 & 0 & D94\\
%-----------------------------------------------------------------------%
 KA~2       & HE 1116$-$1628 & 16.6 & 1.3 & $-$\decsec{0}{21} & \decsec{0}{24}
                                                    & 1 & 0 & 1 & D94\\
%-----------------------------------------------------------------------%
 G77-61     & HE 0330$+$0148 &  15.0 & 1.4 & \decsec{1}{9} & $-$\decsec{7}{5}
                                                    & 0 & 0 & 0 & D94\\\hline\hline
%-----------------------------------------------------------------------%
    \end{tabular}
  \end{center}
\end{table*}

In our test, \emph{all} 21 stars not known as dwarfs were selected either by
their strong C$_2$ bands, or their CN bands. The simulated spectra were also
\emph{all} selected. Of the three dCs, one (KA~2) was selected, and the other
two (G77--61, LHS~1075) not.  From these results we conclude that our sample
of giant C stars and dwarfs with low p.m. (e.g. dCs belonging to the disk
population) is highly complete.  From the small number of dCs in our test
sample we are not able to draw any definitive conclusions, but our results
suggest that only a minor fraction of the halo dCs are detected in the
HES. This is consistent with $\sim 20$\,\% of the simulated halo dCs being found
(see Sect. \ref{sect:dC_selfunct}).

\section{The Surface Density of C Stars}

In an effective area of $6\,400\,\mbox{deg}^2$ \citep[329 of 380 the HES
plates; for a description how the effective area is estimated
  see][]{hespaperIII}, we have isolated a total of 403 C stars. A
straightforward estimate of the surface density of FHLC stars we detect with
the HES is hence obtained from the ratio of these two numbers, yielding
$0.063$\,deg$^{-2}$.  However, one has to take into account that the effective
area accessible on average for each object depends on its brightness,
since the HES limiting magnitude varies from plate to plate. For example,
  the effective area for an object as faint as $B=17.0$ is only $73$\,\% of
  the overall survey area, mainly because only 254 of the contributing plates
  reach this magnitude \citep[see Fig. 2 in][]{hespaperIII}. An additional
  brightness dependence is caused by the fact that faint objects are more
easily affected by overlapping spectra than bright objects. We therefore
determine the FHLC surface density as follows:
\begin{equation}
 \mbox{surface density}=\sum_{i=1}^{403}\frac{1}{\mbox{effarea}(B_J)_i},  
\end{equation}
where $\mbox{effarea}(B_J)$ was determined as described in \cite{hespaperIII}.
We obtain a FHLC surface density of $0.072\pm 0.005$\,deg$^{-2}$ on the 329
HES plates we used.

\section{Discussion and Conclusions}\label{conclusions}

In an effective area of $6\,400\,\mbox{deg}^2$ we have isolated a total
of 403 C stars. Our efforts have thus already increased the number of known
FHLC stars by a factor of nearly five.

We find almost quadruple the surface density of carbon stars compared to the
surveys summarized by \cite{Greenetal:1994}.  Since those previous surveys
claimed limiting magnitudes about $1.5$\,mag fainter than the HES, this
highlights the greatly enhanced selection sensitivity of FHLC stars in the
HES, which is more sensitive to a variety of C$_2$ or CN molecular absorption
band strengths.  Automated selection techniques may be superior to visible
inspection of objective-prism spectra with binocular microscopes, as done e.g.
in the survey of \cite{Sanduleak/Pesch:1988}. Photometric surveys for C stars
have generally selected red objects only, which preferentially selects mostly
the much less common high latitude AGB stars.  \cite{Margonetal:2000} report a
FHLC star surface density of ``at least'' $0.04$\,deg$^{-2}$ in the SDSS. This
is still almost a factor of 2 below our value, and again, the SDSS is much
deeper than the HES ($r'<19.5$).

Due to an average epoch difference of $13.5$ years between DSS~I and HES
plates, we expect to detect and select only $\sim 20$\,\% of the halo dCs that
we could detect and select if direct plates had been taken simultaneously with
the HES plates. Our simulations indicate that 10--15 out of the 403 FHLCs
published in this paper are dCs. Note, however, that this number is uncertain,
because the kinematics of halo dCs is not precisely known. We estimate that
additional $\sim 40$ dCs are detectable on the HES plates, but are
currently missed due to the epoch difference problem. We are extending the
current sample to include proper-motion corrected input catalogs for the
extraction of HES spectra, to find \emph{all} dCs, and other objects that can
have large proper motions, like halo white dwarfs.

\begin{acknowledgements}
  We thank D. Koester for providing model spectra of DQs, and C. Fechner for
  technical support in preparing this article. This work was partly supported
  by Deutsche Forschungsgemeinschaft under grant Re~353/40. P.J.G.
  acknowledges support through NASA Contract NAS8-39073 (ASC).
\end{acknowledgements}

\bibliography{carbonstars,datanaly,imageprocessing,mphs,ncastro,ncpublications,quasar,HES}
\bibliographystyle{apj}

\begin{appendix}

  \section{HES example spectra of C stars}\label{sect:Cstarexamples}

  \begin{figure*}[H]
    \begin{center}
      \leavevmode
      \epsfig{file=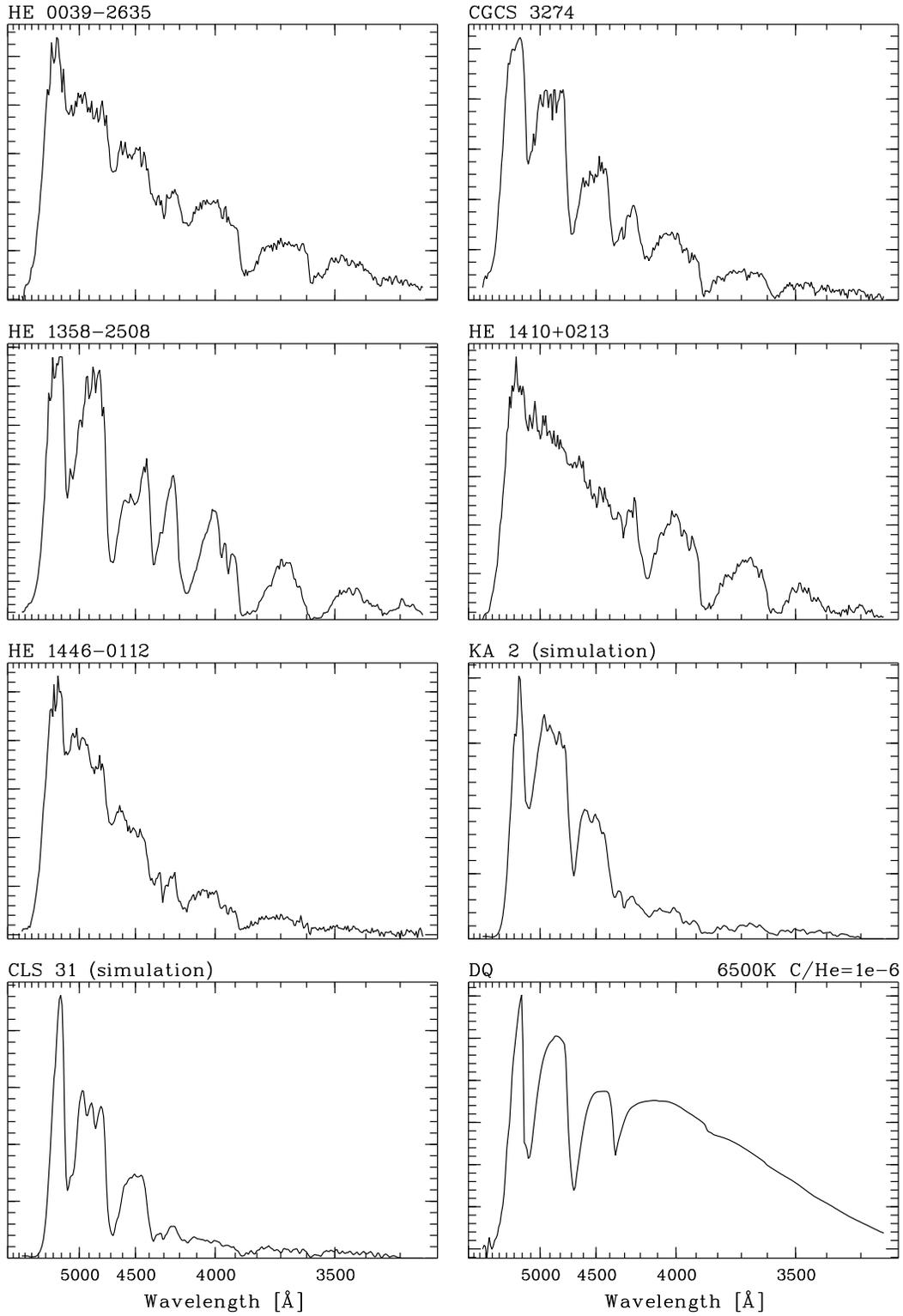, clip=, width=14cm,
        bbllx=80, bblly=125, bburx=497, bbury=736}
      \caption{\label{Cstar_templates} HES spectra of a representative sample of
        seven C stars, displaying a variety of C$_2$ and CN band strengths. The
        spectra of KA~2 and CLS~31 were converted to objective-prism spectra
        from slit spectra with the procedures described in Paper~I. For
        comparison, the spectrum of a DQ white dwarf with
        $T_{\mbox{\scriptsize eff}}=6\,500$\,K and
        $\mbox{C}/\mbox{He}=10^{-6}$ is shown in the lower right panel. That
        star has $U-B=-0.6$.
        }  
    \end{center}
  \end{figure*}

  \section{The HES FHLC sample}

  In Tab. B.1 we list the sample of 403 HES FHLC stars described in this
  paper. The table is made available only electronically. It contains the
  following columns:

  \begin{flushleft}
  \begin{tabular}{ll}
    hename   & HE designation\\
    ra2000   & R.A. at equinox 2000.0, derived from DSS~I\\
    dec2000  & Declination at equinox 2000.0, derived from DSS~I\\
    field    & ESO-SERC field number\\
    plate    & HES plate number\\
    q        & Plate quarter\\
    objtyp   & Object type (stars/bright/ext)\\
    B\_J      & $B_J$ magnitude\\ 
    $V$      & $V$ magnitude\\ 
    $B-V$    & $B-V$ magnitude, derived from HES spectra\\
    $U-B$    & $U-B$ magnitude, derived from HES spectra\\
    C2idx1   & Band index of C$_2$ 5165\,{\AA}\\
    C2idx2   & Band index of C$_2$ 4737\,{\AA}\\
    CNidx1   & Band index of CN 4216\,{\AA}\\
    CNidx3   & Band index of CN 3883\,{\AA}\\
    selC2    & C$_2$ band index selection flag\\
    selCN    & CN band index selection flag\\
  \end{tabular}
  \end{flushleft}
  $B_J$ magnitudes are accurate to better than $\pm 0.2$\,mag, including zero
  point errors \citep{hespaperIII}. $V$ magnitudes were derived by the
  procedures described in Paper~I.  The object types ``stars'',
  ``bright'' and ``ext'' refer to point sources, sources above a saturation
  threshold, and sources detected as extended in DSS~I images, respectively.
  We do not list $V$, $B-V$ and $U-B$ for saturated objects, because our color
  calibrations are not valid for them.
  
\end{appendix}

\end{document}